\documentclass{article}
\usepackage[final, nonatbib]{nips_2017}
\usepackage{amsmath,amssymb,graphicx,multirow,listings,hyperref}
\usepackage{comment}

\newcommand{\sboxed}[1]{\textbf{#1}}

\newcommand{\thet}[1]{\theta_{\sboxed{#1}}}
\newcommand{\ft}{\left(f,t\right)}
\newcommand{\ftt}[1]{\left(f,t\mid\thet{#1}\right)}

\begin{document}
\title{The 2018 Signal Separation Evaluation Campaign}

\author{
  Fabian-Robert St\"oter \\
  Inria and LIRMM, University of Montpellier, France
  \And
  Antoine Liutkus \\
  Inria and LIRMM, University of Montpellier, France
  \And
  Nobutaka Ito \\
  NTT Communication Science Laboratories, NTT Corporation, Japan
}

\maketitle
\begin{abstract}
This paper reports the organization and results for the 2018 community-based Signal Separation Evaluation Campaign (SiSEC 2018). This year's edition was focused on audio and pursued the effort towards scaling up and making it easier to prototype audio separation software in an era of machine-learning based systems. For this purpose, we prepared a new music separation database: MUSDB18, featuring close to 10~h of audio. Additionally, open-source software was released to automatically load, process and report performance on MUSDB18. Furthermore, a new official Python version for the \texttt{BSS~Eval} toolbox was released, along with reference implementations for three oracle separation methods: ideal binary mask, ideal ratio mask, and multichannel Wiener filter. We finally report the results obtained by the participants.
\end{abstract}

\setcounter{footnote}{0}
\section{Introduction}
Source separation is a signal processing problem that consists in recovering individual superimposed \textit{sources} from a \textit{mixture}.  Since 2008, the role of the Signal Separation Evaluation Campaign (SiSEC) has been to compare performance of separation systems on a voluntary and community-based basis, by defining tasks, datasets and metrics to evaluate methods~\cite{sassec2007,sisec2008,sisec0710,sisec2011,sisec2013,sisec2015,sisec2016}. Although source separation may find applications in several domains, the focus of SiSEC has always mostly been on audio source separation.

This year, we decided to drop the legacy speech separation and denoising tasks UND and BGN, because they are now the core focus of very large and successful other campaigns such as CHiME~\cite{chime,chime2,chime3}. Instead, most of our efforts were spent on music separation, where the SiSEC MUS task is playing an important role, both in terms of datasets and participation. However, we also maintained the ASY task of asynchronous separation, due to its originality and adequation with the objectives of SiSEC.

While the primary objective of SiSEC is to regularly report on the progress made by the community through standardized evaluations, its secondary objective is also to provide useful resources for research in source separation, even outside the scope of the campaign itself. This explains why the SiSEC data has always been made public, to be used for related publications.

Since 2015, the scope of the SiSEC MUS data was significantly widened, so that it could serve not only for evaluation, but also for the design of music separation system. This important shift is motivated by the recent development of systems based on deep learning, which now define the state-of-the-art and require important amounts of learning data. This lead to the proposal of the MSD~\cite{sisec2015} and the DSD100~\cite{sisec2016} datasets in the previous editions.

This year's SiSEC present several contributions. First, the computation of oracle performance goes further than the usual Ideal Binary Mask (IBM) to also include Ideal Ratio Mask (IRM) and Multichannel Wiener Filters (MWF). Second, we released the MUSDB18, that comprises almost 10~h of music with separated stems. Third, we released a new version~4 for the \texttt{BSS~Eval} toolbox, that handles time-invariant distortion filters, significantly speeding up computations\footnote{\url{sisec.inria.fr}.}.

\section{Oracle performance for audio separation}
\label{sec:oracle}

We write $I$ as he number of channels of the audio mixture: $I=2$ for stereo. We write $x$ for the 3-dimensional complex array obtained by stacking the Short-Time Frequency Transforms (STFT) of all channels. Its dimensions are $F\times T\times I$, where $F,T$ stand for the number of frequency bands and time frames, respectively. Its values at Time-Frequency (TF) bin $\ft$ are  written $x\ft\in\mathbb{C}^I$, with entries $x_i\ft$. The mixture is the sum of the sources \textit{images}: $x\ft=\sum_j y_j\ft$, which are also multichannel.

A filtering method $\sboxed{m}$ usually computes estimates $\hat{y}_j^{\sboxed{m}}$ for the source images linearly from $x$:
\begin{equation}
  \hat{y}_j^{\sboxed{m}}\ftt{m}=M_j^{\sboxed{m}}\ftt{m} x\ft,\label{eq:TFmask}
\end{equation}
where $\thet{m}$ are some parameters specific to $\sboxed{m}$ and $M_j\ftt{m}$ is a $I\times I$ complex matrix called a TF \textit{mask}, computed using $\thet{m}$ in a way specific to method~$\sboxed{m}$. Once given the filtering strategy $\sboxed{m}$, the objective of a source separation system is to analyze the mixture to obtain parameters $\thet{m}$ that yield good separation performance.

For evaluation purposes, it is useful to know how good a filtering strategy can be, i.e. to have some upper bound on its performance, which is what an \textit{oracle} is~\cite{vincent2007oracle}:
\begin{equation}
  \thet{m}^{\star}=\underset{\thet{m}}{\text{argmin}}\sum_{f,t,j}\left\Vert y_{j}\ft-\hat{y}_{j}^{\sboxed{m}}\ftt{m}\right\Vert,
  \end{equation}
where $\Vert\cdot\Vert$ is any norm deemed appropriate. In this SiSEC, we covered the three most commonly used filtering strategies, and assessed performance of their respective oracles:
\begin{enumerate}
  \item The \textbf{Ideal Binary Mask} (\textit{IBM},~\cite{wang2005}) is arguably the simplest filtering method. It processes all $\left(f,t,i\right)$ of the mixture independently and simply assigns each of them to one source only:   $M_{ij}^\sboxed{IBM}\ft\in\left\{0,1\right\}$. The IMB1 method is defined as $M_{ij}=1$ iff source $j$ has a magnitude $\left|y_{ij}(f,t)\right|$ that is at least half the sum of all sources magnitudes. IBM2 is defined similarly with the sources power spectrograms $\left|y_{ij}(f,t)\right|^2$.
  \item The \textbf{Ideal Ratio Mask} (IRM), also called the $\alpha$-Wiener filter~\cite{liutkus15}, relaxes the binary nature of the IBM. It processes all $\left(f,t,i\right)$ through multiplication by $M_{ij}^\sboxed{IRM}\in\left[0,1\right]$ defined as:
  \begin{equation}
    M^{\sboxed{IRM}}_{ij}\ft=\frac{v_{ij}\ft}{\sum_{j'}v_{ij'}\ft},
  \end{equation}
where $v_{ij}\ft=\left|y_{ij}\ft\right|^\alpha$ is the fractional power spectrogram of the source image $y_{ij}$. Particular cases include the \textit{IRM2} Wiener filter for $\alpha=2$ and the \textit{IRM1} magnitude ratio mask for $\alpha=1$.
  \item The \textbf{Multichannel Wiener Filter} (\textit{MWF}, \cite{duong10}) exploits multichannel information, while IBM and IRM do not. $M^{\sboxed{MWF}}_{j}\ft$ is a $I\times I$ complex matrix given by:
  \begin{equation}
    M_{j}^{\sboxed{MWF}}\ft=C_{j}\ft C_{x}^{-1}\ft,
  \end{equation}
where $C_j\ft$ is the $I\times I$ covariance matrix for source $j$ at TF bin $\ft$ and $C_x=\sum_j C_j$. In the classical local Gaussian model \cite{duong10}, the further parameterization $C_j\ft=v_j\ft R_j\left(f\right)$ is picked, with $R_j$ being the $I\times I$ \textit{spatial covariance matrix}, encoding the average correlations between channels at frequency bin $f$, and $v_j\ft\geq0$ encoding the power spectral density at $\ft$. The optimal values for these parameters are easily computed from the true sources $y_j$ \cite{liutkus2013}.
\end{enumerate}

These five oracle systems IBM1, IBM2, IRM1, IRM2, MWF have been implemented in Python and released in an open-source license\footnote{\url{github.com/sigsep/sigsep-mus-oracle}}.

\section{Data and metrics}

\subsection{The MUSDB18 Dataset}
For the organization of the present SiSEC, the MUSDB18 corpus was released~\cite{musdb18}, comprising tracks from MedleyDB~\cite{medleydb}, DSD100~\cite{sisec2015,sisec2016}, and other material. It contains $150$ full-length tracks, totaling approximately~$10$~h of audio.
\begin{itemize}
\item All items are full-length tracks, enabling the handling of long-term musical structures, and the evaluation of quality over silent regions for sources.
\item All signals are stereo and mixed using professional digital audio workstations, thus representative of real application scenarios.
\item All signals are split into 4 predefined categories: bass, drums, vocals, and other. This promotes automation of the algorithms.
\item Many musical genres are represented: jazz, electro, metal, etc.
\item It is split into a training (100 tracks, 6.5~h) and a test set (50 tracks, 3.5~h), for the design of data-driven methods.
\end{itemize}
The dataset is freely available online, along with Python development tools\footnote{\url{https://sigsep.github.io/musdb}}.

\subsection{BSS Eval version~4}
\label{ssec:bssevalv4}

The BSS~Eval metrics, as implemented in the MATLAB toolboxes~\cite{bssevalv2,bssevalv3} are widely used in the separation literature. They assess separation quality  through $3$~criteria: Source to Distortion, to Artefact, to Interference ratios (SDR, SAR, SIR) and additionally with the Image to Spatial distortion (ISR) for the \texttt{BSS~Eval v3} toolbox~\cite{bssevalv3}.

One particularity of BSS~Eval is to compute the metrics after optimally matching the estimates to the true sources through linear \textit{distortion filters}. This provides some robustness to linear mismatches. This matching is the reason for most of the computation cost of BSS~Eval, especially considering it is done for each evaluation window.

In this SiSEC, we decided to drop the assumption that distortion filters could be varying over time, but considered instead they are fixed for the whole length of the track. First, this significantly reduces the computational cost because matching is done only once for the whole signal. Second, this introduces more dynamics in the evaluation, because time-varying matching filters over-estimate performance, as we show later. Third, this makes matching more stable, because sources are never silent throughout the whole recording, while they often were for short windows.

This new $4^{th}$ version for the \texttt{BSS~Eval} toolbox was implemented in Python\footnote{\texttt{pip install museval}}, and is fully compatible with earlier MATLAB-based versions up to a tolerance of $10^{-12}$~dB in case time-varying filters are selected.

\section{Separation results}
\subsection{Oracle performance with \texttt{BSS Eval v4}}
\label{ssec:bsseval-results}

To the best of our knowledge, the results presented in Figure~\ref{fig:boxplots_bsseval} are the first fair comparison between the different and widely used oracle systems presented in Section~\ref{sec:oracle}. On this figure, we can see boxplots of the \texttt{BSS~Eval} scores obtained by IBM1, IBM2, IRM1, IRM2 and MWF on the $4$~sources considered in MUSDB18. The scores were computed on 1~second windows, taken on the whole test-set.

The most striking fact we see on this Figure~\ref{fig:boxplots_bsseval} is that IBM is \textit{not} achieving the best scores on any metric except ISR. Most particularly, we notice that IBM systematically induces a small loss in performance of a few dBs on SDR and SIR compared to soft masks for most sources, and to a significant loss for SAR, that can get as bad as around~$5$~dB for the accompaniment source. This is in line with the presence of strong \textit{musical noise} produced by IBM whenever the source to separate is \textit{dense} and cannot be assumed stronger in magnitude or energy than all others whenever it is active. This also happens for the bass, which is usually weaker than all other sources at high frequencies, yielding significant distortion with IBM. Furthermore, we suspect the strong scores obtained by IBM in vocals and bass ISR to mostly be due to the zeroing of large amounts of frequency bands in those estimates. Indeed, zero estimates lead the projection filters of BSS~eval to totally cancel those frequencies in the reference also, artificially boosting ISR performance.

Now, comparing soft masks, it appears that IRM2 and MWF produce the best overall performance as compared to IRM1. However, this result is expected: \texttt{BSS~Eval} scores are \textit{in fine} relative to squared-error criteria, which are precisely optimised with those filters. Previous perceptual studies showed that IRM1 may be preferred in some cases~\cite{liutkus15}. This may be reflected in the slightly better performance that IRM1 obtains for SAR. Finally, although IRM2 seems slightly better than MWF for most metrics, we highlight that it also comes with twice as many parameters: power spectral densities for left and right channels, instead of just one for MWF, shared across channels.

\begin{figure}[ht]
  \begin{center}
     \includegraphics[width=0.7\linewidth]{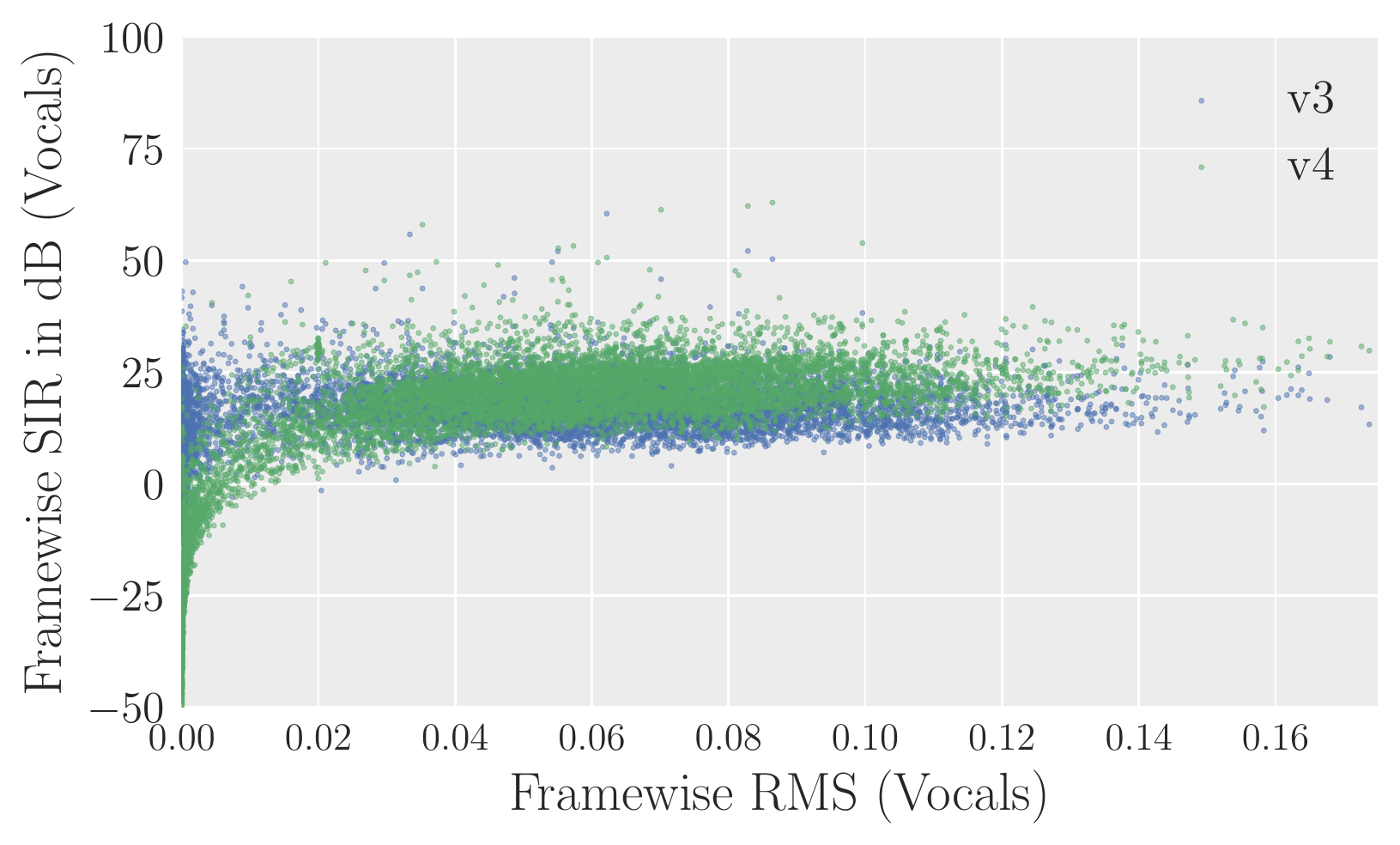}
  \end{center}
  \caption{Vocals SIR score vs vocals energy for BSS~eval v3 and v4.}%
\label{fig:v3v4}
\end{figure}

Concerning the discrepancies between\texttt{BSS~Eval} v3 and v4 (time-invariant distortion filters), we observe several differences. First, computations were $8$~times faster for v4 than for v3, which allowed using small $1$~s frames and thus get an estimate of the performance along time at a reasonable computing cost. Second, computing distortion filters only once for the whole duration of the signal brings an interesting side-effect, that can be visualized on Figure~~\ref{fig:v3v4}. The new v4 brings a much higher dynamics for the scores: we clearly see that lower energy for the true source brings lower performance. However, the marginal distributions for the scores over the whole dataset were not statistically different between v3 and v4, which validates the use of fewer distortion filters to optimize computing time and get to similar conclusions.

\subsection{Comparison of systems submitted to SiSEC-MUS 2018}
This year's participation has been the strongest ever observed for SiSEC, with $30$ systems submitted in total. Due to space constraints, we cannot detail all the methods here, but refer the interested reader to the corresponding papers. We may distinguish three broad groups of methods, that are:
\begin{description}
  \item[Model-based] These methods exploit prior knowledge about the spectrograms of the sources to separate and do not use the MUSDB18 training data for their design. They are: MELO as described in \cite{MELO}, as well as all the method implemented in NUSSL \cite{NUSSL}: 2DFT \cite{2DFT}, RPCA \cite{RPCA}, REP1 \cite{REP1}, REP2 \cite{REP2}, HPSS \cite{HPSS}.
  \item[No additional data] These methods are data-driven and exploit only the
  training data for MUSDB18 to learn the models. They are: RGT1-2 \cite{RGT1}, STL, HEL1 \cite{HEL1}, MDL1 \cite{MDL1}, MDLT \cite{MDLT}, JY1-3 \cite{JY1}, WK \cite{WK}, UHL1-2 \cite{UHL}, TAK1 \cite{TAK12}.
  \item[With additional data] These methods are also data-driven, and exploit additional training data on top of the MUSDB18 training set. They are: UHL3 \cite{UHL}, TAK2 \cite{TAK12}, TAK3 \cite{TAK3}, TAU \cite{TAK3,UHL}.
\end{description}

As may be seen, the vast majority of methods submitted this year to SiSEC MUS are based on deep learning, reflecting a shift in the community's methodology. The MIX method additionally serves as a negative anchor, that corresponds to using the mixture as an estimate for all sources.
\begin{figure}
  \begin{center}
     \includegraphics[height=\textheight]{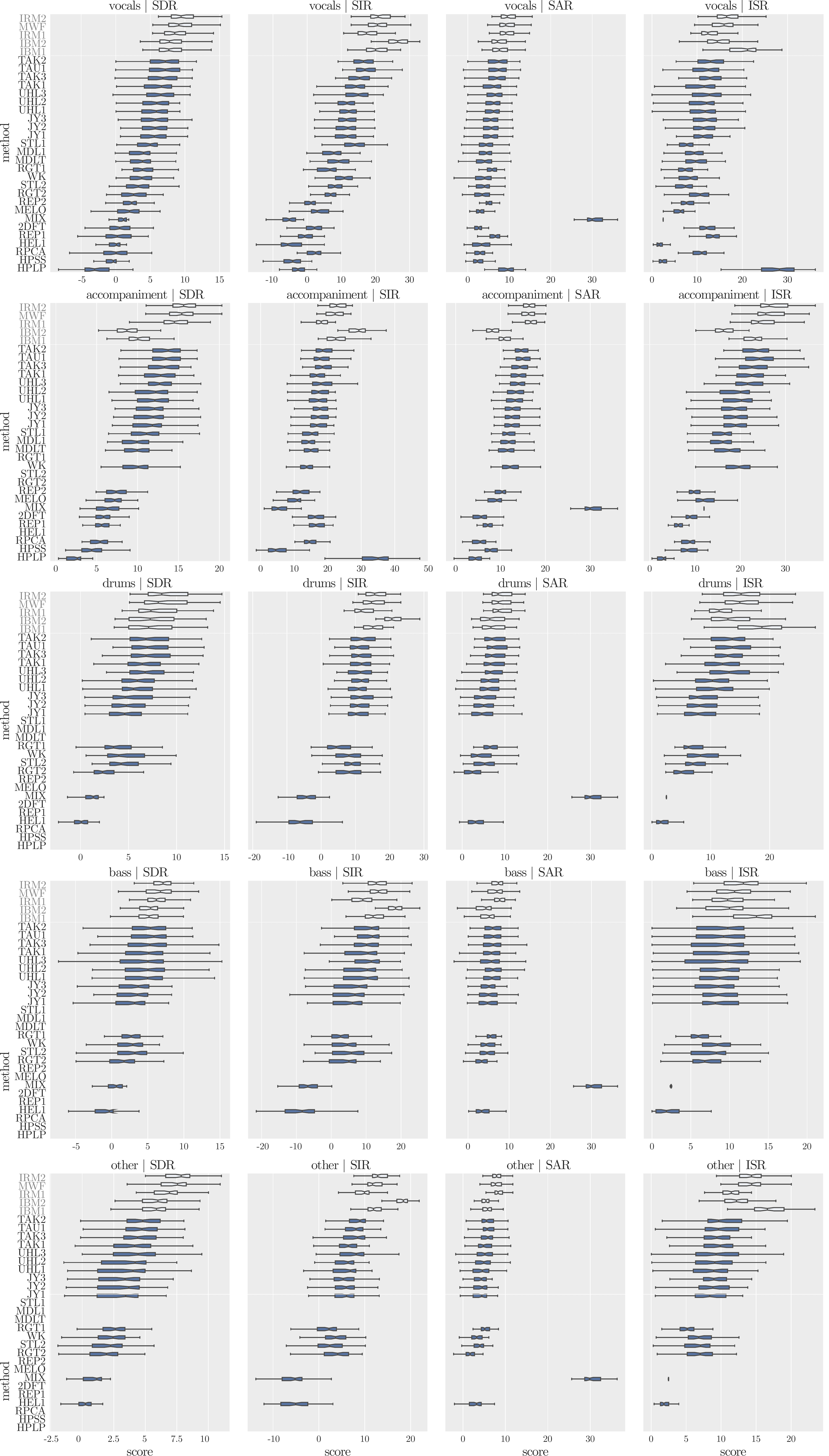}
  \end{center}
  \caption{Details of results for all metrics, targets and methods.}
  \label{fig:boxplots_bsseval}
\end{figure}

In the first set of results depicted on Figure \ref{fig:boxplots_bsseval}, we display boxplots of the BSSeval scores for the evaluation. For each track, the median value of the score was taken and used for the boxplots. Inspecting these results, we immediately see that data-driven methods clearly outperform model-based approaches by a large margin. This fact is noticeable for most targets and metrics.

\begin{figure}
  \begin{center}
     \includegraphics[width=0.85\linewidth]{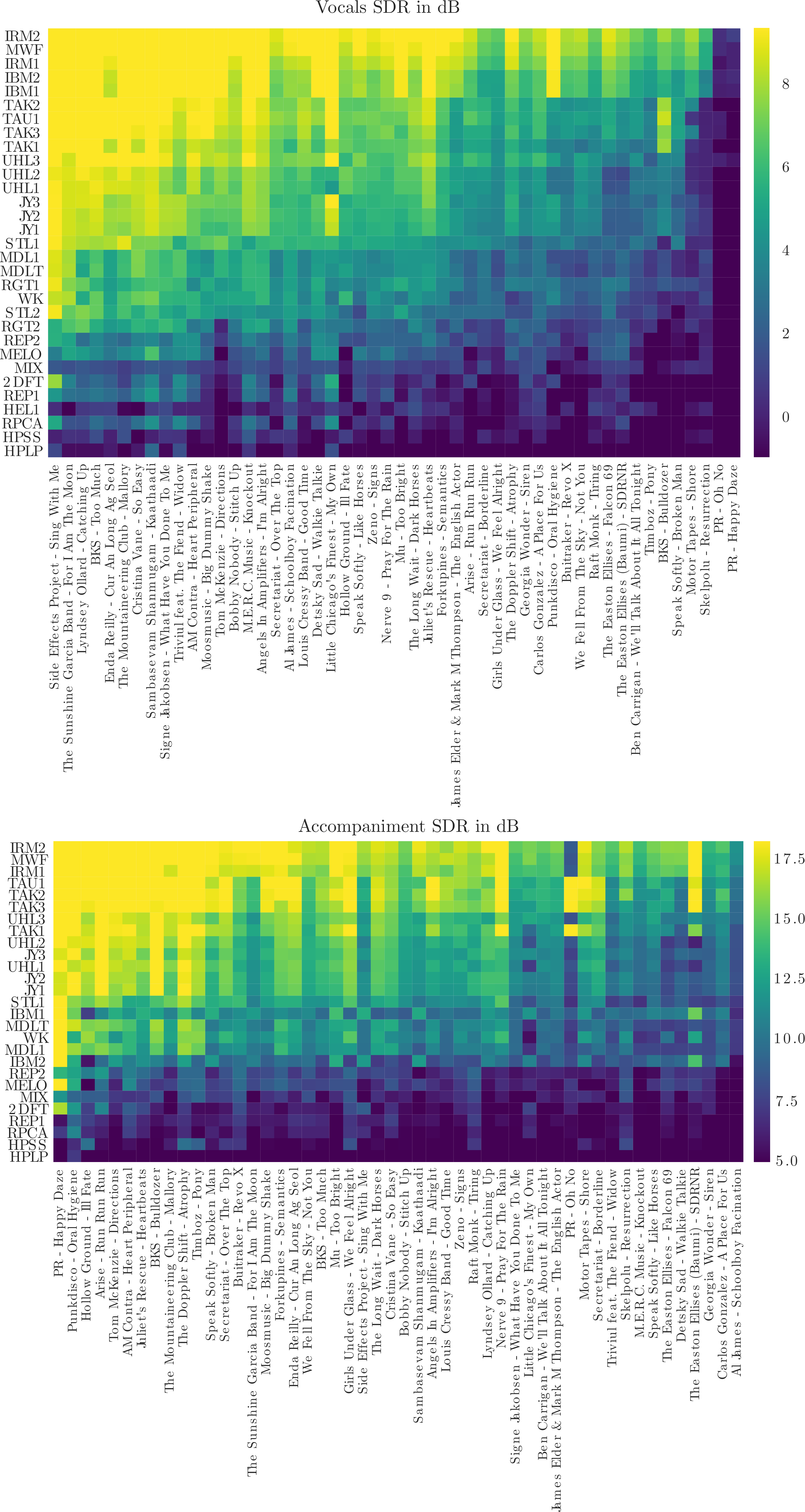}
  \end{center}
  \caption{Vocals (top) and accompaniment (below) SDR for all tracks and methods.}
  \label{fig:trackwise_scores}
\end{figure}

In the second set of results displayed on Figure \ref{fig:trackwise_scores}, we computed the track-wise median SDR score for all methods on the vocals (top) and accompaniment (bottom) targets. The striking fact we notice there is that methods exploiting additional training data (UHL3, TA*) do perform comparably to the oracles for approximately half of the tracks. After inspection, it turns out that room for improvement mostly lies in tracks featuring significant amounts of distortion in either the vocals or the accompaniment. We may also notice on these plots that tracks where accompaniment separation is easy often come with a challenging estimation of vocals. After inspection, this is the case when vocals are rarely active. Consequently, correctly detecting vocals presence seems a good asset for separation methods.

\begin{figure}[h]
  \begin{center}
     \includegraphics[width=1\linewidth]{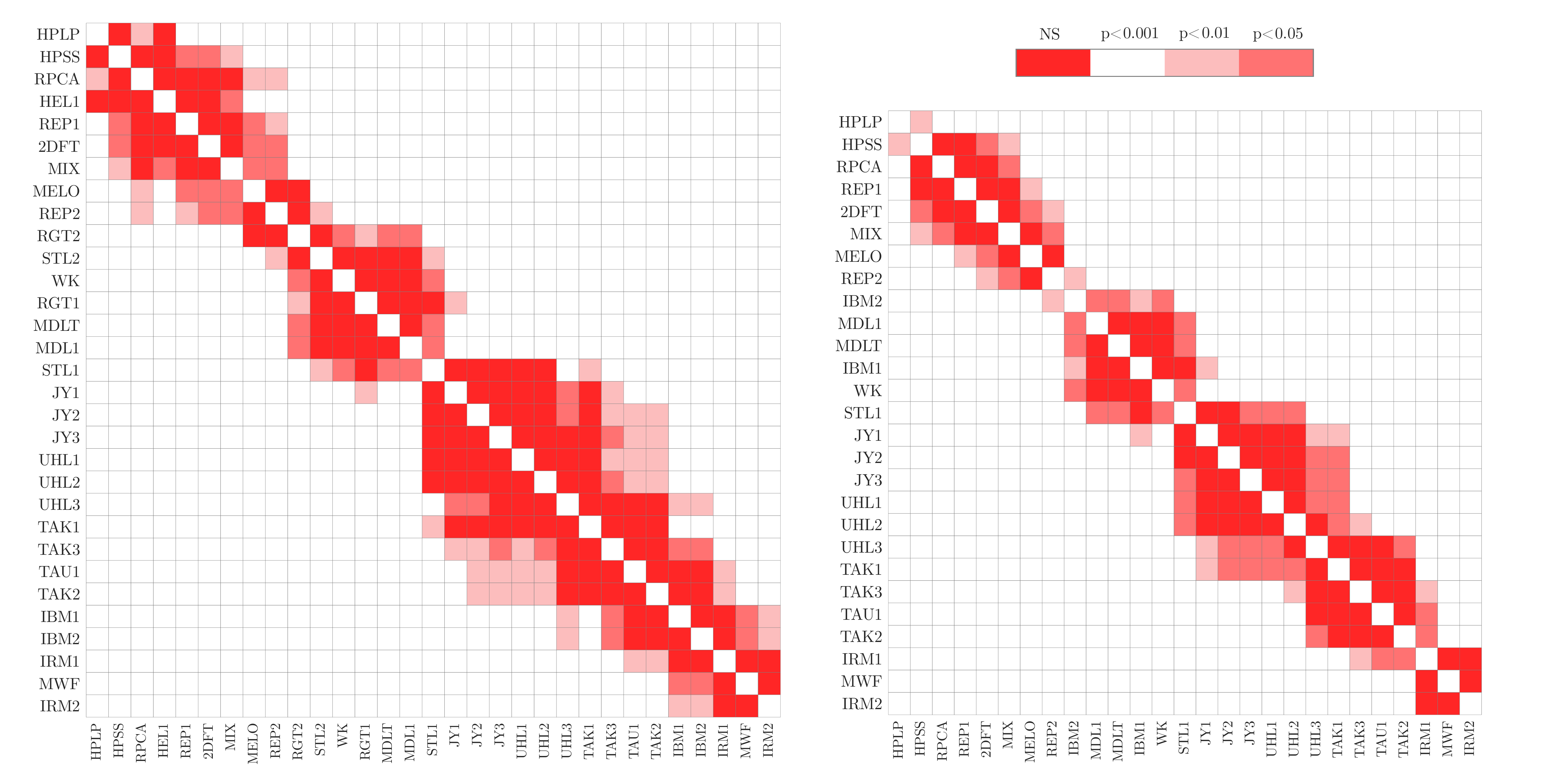}
  \end{center}
  \caption{Pair-wise statistical significance of the differences between separation quality. Left: vocals SDR. Right: accompaniment SDR.}
  \label{fig:pairwise_matrix}
\end{figure}

Our third round of analysis concerns the pair-wise post-hoc Conover-Inman test, displayed on Figure \ref{fig:pairwise_matrix}, to assess which methods perform significantly better than others, for both vocals and accompaniment separation. In this plot, an obvious fact is that DNN-based methods exploiting additional training data perform best. Remarkably, they do not perform significantly differently than the oracles for accompaniment, suggesting that the automatic karaoke problem can now be considered solved to a large extent, given sufficient amounts of training data. On the contrary, vocals separation shows room for improvement.

Concerning model-based methods, we notice they perform worse, but that among them, MELO stands above for vocal separation, while it is comparable to others for accompaniment. For DNN approaches not using additional training data, we notice different behaviours for vocals and accompaniment separation. We may summarize the results by mentioning that RGT1-2, STL and MDL1 do not behave as well as MDLT, STL1, JY1-3, WK and UHL1-2, which all behave comparably. It is noteworthy that TAK1 and UHL2 compare well with methods exploiting additional data for vocals separation.

 This evaluation highlights a methodological question that should be investigated in future campaigns, which is the relative importance of the system architecture and the amount of training data. It indeed appears that very different architectures do behave comparably and that the gap in performance now rather comes from additional training data, as exemplified by the difference between UHL2 and UHL3. This confirms the importance of using standard training and test datasets such as MUSDB18 for evaluation, and we believe that obtaining good performance with reduced training data remains an interesting and challenging machine learning problem.

\subsection{Comparison of systems submitted to SiSEC-ASY 2018}
As shown in Table~\ref{table}, there was one submission to the task "Asynchronous recordings of speech mixtures" by Corey {\it et al.}~\cite{corey}.
This method does not resample the microphone signals in order to separate them. Rather, it uses a separate time-varying two-channel Wiener filter for each synchronous pair of microphones. The remaining asynchronous microphone pairs are used to compute a speech presence probability for each source in each time-frequency bin. The speech presence information from the remote microphone pairs allows the reference recorder to separate more than two speech signals using a two-channel filter.

\begin{table}
\centering
\caption{Result for the task "Asynchronous recordings of speech mixtures". Result by Miyabe {\it et al.} in SiSEC2015 is also shown as a reference.}
\label{table}
\begin{tabular}{|c|c|ccc|ccc|}\hline
systems&criteria&\multicolumn{3}{|c|}{3src}&\multicolumn{3}{|c|}{4src}\\
&&realmix&sumrefs&mix&realmix&sumrefs&mix\\\hline
Corey~\cite{corey}&SDR&$-4.0$&$-4.0$&$-4.1$&$3.1$&$2.9$&$1.7$\\
&ISR&$-0.1$&$-0.1$&$-0.1$&$7.0$&$6.7$&$5.8$\\
&SIR&$-2.2$&$-1.7$&$-1.9$&$5.4$&$5.0$&$2.4$\\
&SAR&$-13.2$&$-13.1$&$-12.4$&$7.9$&$7.8$&$6.1$\\\hline
Miyabe&SDR&6.9&6.8&10.6&4.0&3.8&3.3\\
&ISR&11.2&11.1&15.1&8.8&8.5&7.3\\
&SIR&11.0&10.9&14.9&6.7&6.4&6.0\\
&SAR&11.7&11.6&15.5&7.8&7.6&7.4\\\hline
\end{tabular}
\end{table}

\section{Conclusion}
\label{sec:concl}
We reported our work on the organization of SiSEC 2018, that comprised the development of a new Python version~$4$ for BSS~Eval to assess performance, that is fully compatible with earlier MATLAB versions and additionally allows for time-invariant distortion filters, significantly reducing computational load. Furthermore, we presented the new MUSDB18 dataset, that gathers 150 music tracks with isolated stems, totaling almost $10$~h of music. Finally, we also provide open-source implementations of $3$ popular oracle methods to provide various upper bounds for performance.

Then, we reported the impact of choosing time-invariant distortion filters for BSS~Eval over time-varying ones and quickly summarized the discrepancies in the performance of the proposed oracles methods with BSS~Eval v3 and v4.

Finally, we provided an overall presentation of the scores obtained by the participants to this year's edition. More detailed analysis and sound excerpts can be accessed online on the SiSEC webpage\footnote{\url{sisec18.unmix.app}}.
\footnotesize
\bibliographystyle{plain}
\bibliography{references}

\begin{thebibliography}{10}

\bibitem{sisec2011}
Shoko Araki, Francesco Nesta, Emmanuel Vincent, Zbyn{\v{e}}k Koldovsk{\'y},
  Guido Nolte, Andreas Ziehe, and Alexis Benichoux.
\newblock {\em The 2011 Signal Separation Evaluation Campaign (SiSEC2011): -
  Audio Source Separation -}, pages 414--422.
\newblock 2012.

\bibitem{chime3}
Jon Barker, Ricard Marxer, Emmanuel Vincent, and Shinji Watanabe.
\newblock The third ‘chime’speech separation and recognition challenge:
  Dataset, task and baselines.
\newblock In {\em Automatic Speech Recognition and Understanding (ASRU), 2015
  IEEE Workshop on}, pages 504--511. IEEE, 2015.

\bibitem{chime}
Jon Barker, Emmanuel Vincent, Ning Ma, Heidi Christensen, and Phil Green.
\newblock The pascal chime speech separation and recognition challenge.
\newblock {\em Computer Speech \& Language}, 27(3):621--633, 2013.

\bibitem{medleydb}
Rachel Bittner, Justin Salamon, Mike Tierney, Matthias Mauch, Chris Cannam, ,
  and Juan~P. Bello.
\newblock {MedleyDB}: A multitrack dataset for annotation-intensive mir
  research.
\newblock In {\em 15th International Society for Music Information Retrieval
  Conference}, Taipei, Taiwan, October 2014.

\bibitem{corey}
Ryan~M Corey and Andrew~C Singer.
\newblock Underdetermined methods for multichannel audio enhancement with
  partial preservation of background sources.
\newblock In {\em IEEE Workshop on Applications of Signal Processing to Audio
  and Acoustics (WASPAA)}, pages 26--30, 2017.

\bibitem{duong10}
Ngoc Q.~K. Duong, Emmanuel Vincent, and R\'{e}mi Gribonval.
\newblock Under-determined reverberant audio source separation using a
  full-rank spatial covariance model.
\newblock {\em IEEE Transactions on Audio, Speech, and Language Processing},
  18(7):1830--1840, September 2010.

\bibitem{bssevalv2}
C{\'e}dric F{\'e}votte, R{\'e}mi Gribonval, and Emmanuel Vincent.
\newblock Bss\_eval toolbox user guide--revision 2.0.
\newblock 2005.

\bibitem{HPSS}
Derry Fitzgerald.
\newblock Harmonic/percussive separation using median filtering.
\newblock 2010.

\bibitem{RPCA}
Po-Sen Huang, Scott~Deeann Chen, Paris Smaragdis, and Mark Hasegawa-Johnson.
\newblock Singing-voice separation from monaural recordings using robust
  principal component analysis.
\newblock In {\em Acoustics, Speech and Signal Processing (ICASSP), 2012 IEEE
  International Conference on}, pages 57--60. IEEE, 2012.

\bibitem{HEL1}
Po-Sen Huang, Minje Kim, Mark Hasegawa-Johnson, and Paris Smaragdis.
\newblock Singing-voice separation from monaural recordings using deep
  recurrent neural networks.
\newblock In {\em ISMIR}, pages 477--482, 2014.

\bibitem{JY1}
Jen-Yu Liu and Yi-Hsuan Yang.
\newblock {JY Music Source Separtion submission for SiSEC, Research Center for
  IT Innovation, Academia Sinica, Taiwan}.
\newblock \url{https://github.com/ciaua/MusicSourceSeparation}, 2018.

\bibitem{liutkus15}
Antoine Liutkus and Roland Badeau.
\newblock Generalized {W}iener filtering with fractional power spectrograms.
\newblock In {\em IEEE International Conference on Acoustics, Speech and Signal
  Processing}, Brisbane, QLD, Australia, April 2015.

\bibitem{liutkus2013}
Antoine Liutkus, Roland Badeau, and Ga{\"e}l Richard.
\newblock Low bitrate informed source separation of realistic mixtures.
\newblock In {\em Acoustics, Speech and Signal Processing (ICASSP), 2013 IEEE
  International Conference on}, pages 66--70. IEEE, 2013.

\bibitem{sisec2016}
Antoine Liutkus, Fabian-Robert St{\"o}ter, Zafar Rafii, Daichi Kitamura,
  Bertrand Rivet, Nobutaka Ito, Nobutaka Ono, and Julie Fontecave.
\newblock The 2016 signal separation evaluation campaign.
\newblock In {\em International Conference on Latent Variable Analysis and
  Signal Separation}, pages 323--332. Springer, 2017.

\bibitem{NUSSL}
Ethan Manilow, Prem Seetharaman, Fatemah Pishdadian, and Bryan Pardo.
\newblock {NUSSL}: the northwestern university source separation library.
\newblock \url{https://github.com/interactiveaudiolab/nussl}, 2018.

\bibitem{MDLT}
Stylianos~Ioannis Mimilakis, Konstantinos Drossos, Joao Santosand~Gerald
  Schuller, Tuomas Virtanen, and Yoshua Bengio.
\newblock Monaural singing voice separation with skip-filtering connections and
  recurrent inference of time-frequency mask.
\newblock 2017.

\bibitem{MDL1}
Stylianos~Ioannis Mimilakis, Konstantinos Drossos, Tuomas Virtanen, and Gerald
  Schuller.
\newblock A recurrent encoder-decoder approach with skip-filtering connections
  for monaural singing voice separation.
\newblock 2017.

\bibitem{sisec2013}
Nobutaka Ono, Zbyn{\v{e}}k Koldovsk{\'y}, Shigeki Miyabe, and Nobutaka Ito.
\newblock The 2013 signal separation evaluation campaign.
\newblock In {\em 2013 IEEE International Workshop on Machine Learning for
  Signal Processing (MLSP)}, Sept 2013.

\bibitem{sisec2015}
Nobutaka Ono, Zafar Rafii, Daichi Kitamura, Nobutaka Ito, and Antoine Liutkus.
\newblock The 2015 signal separation evaluation campaign.
\newblock In {\em International Conference on Latent Variable Analysis and
  Signal Separation}, pages 387--395. Springer, 2015.

\bibitem{REP2}
Zafar Rafii, Antoine Liutkus, and Bryan Pardo.
\newblock Repet for background/foreground separation in audio.
\newblock In {\em Blind Source Separation}, pages 395--411. Springer, 2014.

\bibitem{musdb18}
Zafar Rafii, Antoine Liutkus, Fabian-Robert Stöter, Stylianos~Ioannis
  Mimilakis, and Rachel Bittner.
\newblock The {MUSDB18} corpus for music separation, December 2017.

\bibitem{REP1}
Zafar Rafii and Bryan Pardo.
\newblock Repeating pattern extraction technique (repet): A simple method for
  music/voice separation.
\newblock {\em IEEE transactions on audio, speech, and language processing},
  21(1):73--84, 2013.

\bibitem{RGT1}
Gerard Roma, Owen Green, and Pierre-Alexandre Tremblay.
\newblock Improving single-network single-channel separation of musical audio
  with convolutional layers.
\newblock In {\em International Conference on Latent Variable Analysis and
  Signal Separation}, 2018.

\bibitem{MELO}
Justin Salamon and Emilia G{\'o}mez.
\newblock Melody extraction from polyphonic music signals using pitch contour
  characteristics.
\newblock {\em IEEE Transactions on Audio, Speech, and Language Processing},
  20(6):1759--1770, 2012.

\bibitem{2DFT}
Prem Seetharaman, Fatemeh Pishdadian, and Bryan Pardo.
\newblock Music/voice separation using the 2d fourier transform.
\newblock In {\em Applications of Signal Processing to Audio and Acoustics
  (WASPAA), 2017 IEEE Workshop on}, pages 36--40. IEEE, 2017.

\bibitem{TAK12}
Naoya {Takahashi}, Nabarun {Goswami}, and Yuki {Mitsufuji}.
\newblock {MMDenseLSTM: An efficient combination of convolutional and recurrent
  neural networks for audio source separation}.
\newblock {\em ArXiv e-prints, arxiv:1805.02410}, May 2018.

\bibitem{TAK3}
Naoya Takahashi and Yuki Mitsufuji.
\newblock {M}ulti-{S}cale multi-band densenets for audio source separation.
\newblock In {\em IEEE Workshop on Applications of Signal Processing to Audio
  and Acoustics (WASPAA)}, pages 21--25. IEEE, 2017.

\bibitem{UHL}
Stefan Uhlich, Marcello Porcu, Franck Giron, Michael Enenkl, Thomas Kemp, Naoya
  Takahashi, and Yuki Mitsufuji.
\newblock Improving music source separation based on deep neural networks
  through data augmentation and network blending.
\newblock In {\em IEEE International Conference on Acoustics, Speech and Signal
  Processing (ICASSP)}, pages 261--265. IEEE, 2017.

\bibitem{sisec2008}
Emmanuel Vincent, Shoko Araki, and Pau Bofill.
\newblock The 2008 signal separation evaluation campaign: A community-based
  approach to large-scale evaluation.
\newblock In {\em International Conference on Independent Component Analysis
  and Signal Separation}, pages 734--741. Springer, 2009.

\bibitem{sisec0710}
Emmanuel Vincent, Shoko Araki, Fabian Theis, Guido Nolte, Pau Bofill, Hiroshi
  Sawada, Alexey Ozerov, Vikrham Gowreesunker, Dominik Lutter, and Ngoc~QK
  Duong.
\newblock The signal separation evaluation campaign (2007--2010): Achievements
  and remaining challenges.
\newblock {\em Signal Processing}, 92(8):1928--1936, 2012.

\bibitem{chime2}
Emmanuel Vincent, Jon Barker, Shinji Watanabe, Jonathan Le~Roux, Francesco
  Nesta, and Marco Matassoni.
\newblock The second ‘chime’speech separation and recognition challenge:
  Datasets, tasks and baselines.
\newblock In {\em Acoustics, Speech and Signal Processing (ICASSP), 2013 IEEE
  International Conference on}, pages 126--130. IEEE, 2013.

\bibitem{bssevalv3}
Emmanuel Vincent, R{\'e}mi Gribonval, and C{\'e}dric F{\'e}votte.
\newblock Performance measurement in blind audio source separation.
\newblock {\em IEEE transactions on audio, speech, and language processing},
  14(4):1462--1469, 2006.

\bibitem{vincent2007oracle}
Emmanuel Vincent, R{\'e}mi Gribonval, and Mark~D Plumbley.
\newblock Oracle estimators for the benchmarking of source separation
  algorithms.
\newblock {\em Signal Processing}, 87(8):1933--1950, 2007.

\bibitem{sassec2007}
Emmanuel Vincent, Hiroshi Sawada, Pau Bofill, Shoji Makino, and Justinian~P
  Rosca.
\newblock First stereo audio source separation evaluation campaign: data,
  algorithms and results.
\newblock In {\em International Conference on Independent Component Analysis
  and Signal Separation}, pages 552--559. Springer, 2007.

\bibitem{wang2005}
DeLiang Wang.
\newblock On ideal binary mask as the computational goal of auditory scene
  analysis.
\newblock {\em Speech separation by humans and machines}, pages 181--197, 2005.

\bibitem{WK}
Felix Weninger, John~R Hershey, Jonathan Le~Roux, and Bj{\"o}rn Schuller.
\newblock Discriminatively trained recurrent neural networks for single-channel
  speech separation.
\newblock In {\em IEEE Global Conference on Signal and Information Processing
  (GlobalSIP)}, pages 577--581. IEEE, 2014.

\end{thebibliography}
\end{document}